\documentclass[conference,compsoc]{IEEEtran}

\usepackage{graphicx}
\usepackage{booktabs}
\usepackage{tabularx}
\usepackage{array}
\usepackage{makecell}
\usepackage{adjustbox}
\usepackage{cite}

\usepackage{amsmath}
\usepackage{amsfonts}

\usepackage{algorithm}
\usepackage{algorithmic}

\usepackage{url}
\usepackage{xspace}
\usepackage{xcolor}
\usepackage{listings}
\usepackage{enumitem}
\usepackage{float}
\usepackage[most]{tcolorbox}
\usepackage{caption}

\usepackage{pgfplots}
\pgfplotsset{compat=1.17}
\usepackage{subcaption}
\usepgfplotslibrary{groupplots}

\newcommand{\sname}{\textsc{VulSolver}\xspace}

\title{\sname: Vulnerability Detection via LLM-Driven Constraint Solving}

\author{
\IEEEauthorblockN{
Xiang Li$^*$,\;
Yueci Su\textsuperscript{\dag},\;
Jiahao Liu\textsuperscript{\ddag},\;
Zhiwei Lin\textsuperscript{\dag},\;
Yuebing Hou$^*$,\;
Peiming Gao$^*$,\;
Yuanchao Zhang$^*$
}
\IEEEauthorblockA{
$^*$ MYbank, Ant Group \quad
\textsuperscript{\dag} Ant Group \quad
\textsuperscript{\ddag} National University of Singapore
}
}
\usepackage[breaklinks=true,colorlinks=true,
            linkcolor=black,           
            citecolor=blue,            
            urlcolor=blue]{hyperref}   

\begin{document}
\maketitle

\begin{abstract}
Traditional vulnerability detection methods rely heavily on predefined rule matching, which often fails to capture vulnerabilities accurately.
With the rise of large language models (LLMs), leveraging their ability to understand code semantics has emerged as a promising direction for achieving more accurate and efficient vulnerability detection.
However, current LLM-based approaches face significant challenges: instability in model outputs, degraded performance with long context, and hallucination.
As a result, many existing solutions either use LLMs merely to enrich predefined rule sets, thereby keeping the detection process fundamentally rule-based, or over-rely on them, leading to poor robustness.
To address these challenges, we propose a constraint-solving approach powered by LLMs named \sname.
By modeling vulnerability detection as a constraint-solving problem, and by integrating static application security testing (SAST) with the semantic reasoning capabilities of LLMs, our method enables the LLM to act like a professional human security expert.
We assess \sname on the OWASP Benchmark (1,023 labeled samples), achieving 97.85\% accuracy, 97.97\% F1-score, and 100\% recall. Applied to widely-used open-source projects, \sname identified 15 previously unknown high-severity vulnerabilities (CVSS 7.5-9.8), demonstrating its effectiveness in real-world security analysis.
\end{abstract}

\section{Introduction}
\label{sec:introduction}

With the rapid advancement of software development, software vulnerabilities have also increased in both number and complexity~\cite{ruan2024vulzoo,nguyen2022regvd}.
These vulnerabilities have led to significant security incidents, resulting in severe consequences such as data breaches and financial losses.
For instance, Heartbleed~\cite{mitreCVE20140160}, a critical vulnerability in OpenSSL, allowed attackers to read sensitive data directly from the memory of affected servers, compromising millions of systems that relied on OpenSSL for secure communication.
As such, detecting software vulnerabilities is essential to maintaining the security and reliability of modern software systems.

Existing solutions for vulnerability detection can be broadly categorized into two main classes: rule-based approaches~\cite{gao2018cobot,sui2016svf,lee2006rule} and learning-based approaches~\cite{cao2022mvd,cui2020vuldetector,diwan2022vdgraph2vec,lin2019software}.
Rule-based methods rely on security experts to define heuristics or syntactic/semantic patterns that match known vulnerability conditions.
These approaches are effective at detecting well-defined, previously known vulnerabilities with clear patterns.
However, they are heavily dependent on manual effort, making them labor-intensive, time-consuming, and difficult to scale.
To overcome these limitations, learning-based approaches aim to automatically capture implicit and complex vulnerability patterns from large-scale code corpora.
Typically, these methods represent code in diverse structural forms --- such as abstract syntax trees (ASTs) or control flow graphs (CFGs) --- and then establish a mapping between these representations, including vulnerabilities and their label with machine learning techniques.
Although learning-based approaches have shown promising results, their effectiveness is often constrained by the availability of labeled training data, which is limited and costly to obtain.
Moreover, many of these models are tailored to specific programming languages or vulnerability types, limiting their generalizability and applicability across different software ecosystems~\cite{zhang2024prompt}.
Despite these advancements, effective and efficient detection of vulnerabilities in complex systems continues to be a fundamental and unresolved challenge.

Recent advancements in LLMs\cite{ge2023openagi,yao2024survey} have opened up new opportunities for enhancing vulnerability detection.
LLMs exhibit strong capabilities in language understanding, reasoning, and decision-making\cite{wei2022chain,ho2022large}, allowing them to analyze source code with greater nuance and uncover subtle indicators of potential security flaws.
Nevertheless, directly applying LLMs to vulnerability detection remains challenging.
Real-world software systems often comprise large codebases, where vulnerabilities can span multiple methods.
In such cases, the complexity of large codebases, combined with LLMs' limited context capacity and performance degradation on lengthy inputs, hinder effective analysis, making it difficult to reliably detect vulnerabilities.

To bridge this gap, we formulate vulnerability detection as a path-based constraint-solving problem, where the detection process involves solving constraints derived from program execution paths.
A vulnerability is confirmed when its corresponding constraints are satisfied. 
For instance, consider the code snippet in Listing~\ref{runing_example_lst}, which exhibits a potential arbitrary file-read vulnerability.
To verify its presence, we analyze the execution path \textit{doGet} $\rightarrow$ \textit{read} $\rightarrow$ \textit{readFile} $\rightarrow$ \textit{readString}, which represents the primary execution flow.
The objective is to determine whether the first argument passed to \textit{readString}, \textit{Paths.get(path)}, contains the substring "\texttt{..}", which indicates the possibility of arbitrary file read.

We abstract two types of constraints to guarantee the existence of a vulnerability when all constraints are satisfied:
(1) \textit{Transfer Constraints:} These ensure that the sink point is reachable from the source, i.e., there exists a feasible execution path in the program. 
(2) \textit{Trigger Constraints:} These ensure that the input reaching the sink satisfies the conditions necessary for exploitation—e.g., for path traversal vulnerabilities, the path parameter contains "\texttt{..}".
Specifically, transfer constraints capture one-hop connectivity along the call path (i.e., caller–callee relationships), while trigger constraints verify that parameters at each method invocation maintain the exploitable state—e.g., whether the callee's path parameter can contain "\texttt{..}" when the method is called.
Given these constraints, we leverage LLMs to solve them: For each caller-callee relationship in the call path, assess whether the caller’s context enables the callee invocation (i.e., the call is feasible), and determine whether the actual arguments can propagate a payload to the sink and activate the vulnerability.

In this paper, we implement the proposed detection pipeline as \sname, which takes program code as input and outputs whether it contains specific vulnerabilities.
Specifically, \sname first performs a static analysis to extract potentially vulnerable call paths that serve as the analysis backbone, where nodes represent methods and edges capture their invocation relationships, both retaining detailed contextual information.
Next, based on the type of targeted vulnerability, we extract the corresponding constraints.
For transfer constraints, we traverse the call path and obtain them directly.
For trigger constraints, we initialize the analysis based on the vulnerability type and the sink method.
For instance, in the case of an arbitrary file-read vulnerability, we examine whether the method is reachable and whether any arguments that semantically represent a file path could potentially contain "\texttt{..}".



To assess the effectiveness of \sname in vulnerability detection, we evaluate it on the OWASP Benchmark, which contains 1,023 labeled samples spanning command injection, path traversal, and SQL injection vulnerabilities.
We further apply \sname to real-world programs by mining popular open-source projects to identify in-the-wild security risks.
Powered by Kimi-K2, \sname achieves strong results, reaching 97.85\% accuracy, 97.97\% F1-score, and 100\% recall on OWASP, outperforming existing methods.
Beyond benchmarks, our framework also discovered 15 previously unknown high-severity vulnerabilities (CVSS 7.5–9.8) in real projects, underscoring both its detection capability and its practical value in real-world security analysis.
We also perform a series of ablation studies to isolate and quantify how each design choice contributes to \sname’s overall performance.


In the paper, we make the following contributions.
\begin{itemize}[leftmargin=14pt, topsep=2pt, itemsep=0pt]
    \item We are the first to formulate vulnerability detection as a path-based constraint-solving problem, encompassing both transfer and trigger constraints. This formulation is general across different vulnerability types and amenable to automated processing.
    \item We incorporate both main-path information (i.e., methods directly on the source-to-sink call chain) and branch-path information (i.e., methods transitively invoked from the main path), and guide LLMs to systematically solve constraints by analyzing this structured information in a controlled manner.
    \item We implement \sname and conduct comprehensive evaluations on both the OWASP Benchmark and multiple real-world applications, demonstrating that our approach achieves effective and accurate vulnerability detection while consistently outperforming existing solutions. The core implementation along with all experimental artifacts will be made publicly available upon acceptance to facilitate verification, reproducibility, and further research.
\end{itemize}
\lstdefinestyle{javastyle}{
    basicstyle=\ttfamily\footnotesize\color[rgb]{0.2,0.2,0.2},  
    keywordstyle=\color[rgb]{0.09,0.28,0.48}\bfseries,         
    commentstyle=\color[rgb]{0.42,0.42,0.42}\itshape,          
    stringstyle=\color[rgb]{0.62,0.13,0.22},                   
    numberstyle=\tiny\color[rgb]{0.5,0.5,0.5},                
    emph={String,HttpServletRequest,Exception},               
    emphstyle=\color[rgb]{0.11,0.49,0.35}\bfseries,            
    emph={[2]doGet,save,getPath},                             
    emphstyle=[2]\color[rgb]{0.48,0.15,0.45},                 
    emph={[3]fileName,content,request},                      
    emphstyle=[3]\color[rgb]{0.72,0.33,0.08},                
    backgroundcolor=\color{white},
    frame=none,
    framerule=0.8pt,
    rulecolor=\color[rgb]{0.82,0.82,0.82},                    
    numbers=left,
    numbersep=5pt,
    breaklines=true,
    breakatwhitespace=true,
    tabsize=4,
    showstringspaces=false,
    xleftmargin=12pt,
    aboveskip=8pt,
    belowskip=4pt
}
\section{Preliminaries}
\label{preliminaries}

In this section, we begin with a running example to illustrate how human security experts typically conduct code audits and the information required in this process. 
We then examine the current state of SAST and LLM-based approaches, analyzing their respective strengths and limitations, and identifying the gaps that prevent them from achieving the reliability of human expert auditing.
Building on these insights, we formally define vulnerability detection as a constraint-solving problem, thereby reframing the challenge of effective vulnerability detection as one of systematic constraint solving, and propose a core algorithm to solve this constraint system.

\subsection{Motivating Example: Human Code Auditing}
\label{running_example}

Listing~\ref{runing_example_lst} presents a code snippet that performs file reading operations based on user input.
To determine whether an arbitrary file-read vulnerability exists, a human security expert conducts a systematic, step-by-step analysis rather than examining all code at once.

The expert first identifies the execution flow \textit{doGet} $\rightarrow$ \textit{read} $\rightarrow$ \textit{readFile} $\rightarrow$ \textit{readString}, and then analyzes each caller-callee pair progressively along this chain.
At each step, the expert addresses two questions: 
(1) Can the callee be invoked given the caller's execution context? 
(2) When invoked, do the arguments passed to the callee maintain an exploitable state?
For instance, when examining the pair \textit{doGet}$\rightarrow$\textit{read}, the expert determines whether \textit{read} is reachable and whether the \textit{fileName} argument can contain "\texttt{..}", since this parameter may eventually influence the file path used at the sink function.
By propagating this reasoning operation along the entire call chain, the expert can conclude whether the sink method (\textit{readString}) is ultimately reachable with exploitable arguments—thereby confirming or ruling out the vulnerability.

To perform this analysis at each caller-callee pair, the expert needs three types of information.

First, the caller's code, where it invokes the callee must be examined, as this provides the foundation for analyzing the callee's invocation state.

Second, the expert must account for caller-side context beyond the immediate callee: auxiliary methods invoked by the caller can encode semantics that condition the callee’s state.
For example, when \textit{read} calls \textit{readFile}, examining \textit{read} alone is insufficient; one must also recognize that \textit{getPath} validates \textit{fileName}, which determines the argument state observed by \textit{readFile}.

Third, the state of the caller's parameters when it is invoked must be known. These states summarize all prior processing before the caller is invoked, allowing the expert to reason forward without re-examining earlier code.
For instance, when analyzing \textit{readFile} calling \textit{readString}, the expert must know that \textit{readFile}'s parameter \textit{path} has already been filtered to conclude that the arguments to \textit{readString} are also non-exploitable.

In summary, human expert analysis is not a process of directly reading all code and drawing final conclusions in a single pass, but rather an elegant progressive approach.
The expert analyzes methods one by one, deriving critical conclusions at each step (e.g., whether a method's parameters are exploitable when invoked), which inform subsequent analysis.
In conducting further analysis, the expert does not retain all previously examined code in memory; instead, they build upon prior critical conclusions to continue reasoning forward.
Each iteration yields critical conclusions that are propagated forward, culminating in a determination of whether the sink method is reachable with exploitable inputs.

\subsection{SAST Approaches}
\label{sast}

SAST analyzes code paths to identify potential security vulnerabilities.
It takes source code, bytecode, or intermediate representations as input and applies data-flow, control-flow, and propagation analyses to match predefined rules for detecting vulnerabilities~\cite{li2023comparison, wadhams2024barriers, ruan2025accurate, jiang2025fuzzingphpinterpreterdataflow}.
By tracking the propagation of program states across methods and modules, SAST can efficiently highlight candidate paths where tainted data may flow from untrusted sources to sensitive sinks, surfacing potential security risks for further examination.
While SAST achieves promising detection results, it often suffers from high false-positive rates due to its limited ability to reason about deep program semantics.
Most SAST tools rely on taint analysis to track whether one variable’s value is derived from another, but they frequently struggle to capture the semantic transformations that occur during propagation~\cite{mirsky2023vulchecker}.
For instance, conventional SAST tools often find it difficult to determine whether a value has been sanitized or encoded into a safe representation.

Taking the running example in Listing~\ref{runing_example_lst}, SAST can identify the main path and, through taint analysis, determine that the file path being read originates from user input, leading it to flag a potential file read vulnerability.
However, the \textit{getPath} method already filters the input, preventing path traversal from being exploitable in this case.
While SAST can be improved by incorporating sanitization methods or similar rules, such rules are inherently difficult to exhaustively enumerate~\cite{landman2017challenges}.

\begin{lstlisting}[style=javastyle, caption={A Running example that performs file reading operations based on user input}, label=runing_example_lst, float=t]
public void doGet(HttpServletRequest request, HttpServletResponse response) throws IOException {
    String fileName = request.getParameter("fileName");
    String content = read(fileName);
    response.setContentType("text/plain; charset=UTF-8");
    response.getWriter().write(content);
}

public String read(String fileName) throws IOException {
    String path = getPath(fileName);
    return readFile(path);
}

public String getPath(String fileName) {
    if (!fileName.contains("..")) {
        return "/tmp/files/" + fileName;
    } else {
        throw new IllegalArgumentException("Invalid file name");
    }
}

public String readFile(String path) throws IOException {
    return Files.readString(Paths.get(path));
}
\end{lstlisting}

\begin{table*}[t]
\centering
\renewcommand{\arraystretch}{1.35}
\begin{tabularx}{\textwidth}{>{\centering\arraybackslash}m{0.14\textwidth}|>{\raggedright\arraybackslash}m{0.20\textwidth}|X}
\hline
\textbf{Notation} & \textbf{Name} & \textbf{Definition} \\ \hline \hline
$\mathbf{P}$ & Main Path & Ordered sequence $\mathbf{P} = \langle m_1, m_2, \ldots, m_n \rangle$ from source $m_1$ to sink $m_n$. Each $m_i$ is called a main method. \\ \hline

$\mathbf{N}$ & Branch Methods & For each $m_i \in \mathbf{P}$, the set $\mathbf{N}_{i}$ consists of its directly invoked methods, excluding $m_{i+1}$. Each $n_{i,j} \in \mathbf{N}_i$ denotes the $j$-th branch method of $m_i$. \\ \hline

$\mathbf{T}$ & Branch Trees & For each $n_{i,j} \in \mathbf{N}$, the branch tree $t_{i,j}$ includes $n_{i,j}$ and all methods transitively invoked by it: $t_{i,j} = \{n_{i,j}\} \cup \{ u \mid u \text{ is transitively invoked by } n_{i,j} \}$. The set $\mathbf{T}_{i}$ consists of all branch trees rooted at the methods in $\mathbf{N}_i$. \\ \hline

$\mathbf{C}$ & Critical Types & Types that are capable of carrying security-sensitive semantics (such as "path" semantics for path traversal, "SQL" semantics for SQL injection, or "command" semantics for command execution) are critical types, e.g., String and Path are critical types for path traversal vulnerabilities in Java. Each vulnerability category corresponds to an enumerable set of critical types, which are drawn from commonly available types provided by the programming language. \\ \hline

$\mathbf{U}$ & \makecell[l]{Non-exploitable Conditions} & Each vulnerability type associates every critical type $c \in \mathbf{C}$ with a condition under which values of $c$ cannot be exploited. For example, in path traversal vulnerabilities, a variable of the critical type Path cannot be exploited if ".." in its value has been strictly filtered.\\ \hline

$\mathbf{A}$ & Critical Parameters & For each $m_i \in \mathbf{P}$, the set $\mathbf{A}_i$ consists of all parameters of $m_i$ that are critical parameters. A parameter is a critical parameter if its type is in $\mathbf{C}$ (i.e., the parameter itself is of a critical type) or encapsulates a type in $\mathbf{C}$ (i.e., the parameter wraps a critical type). Each $a_{i,k}$ denotes the $k$-th critical parameter of $m_i$. \\ \hline

$\mathbf{S}$ & Parameter States & For each $m_i \in \mathbf{P}$, the set $\mathbf{S}_i$ consists of the states of all critical parameters in $\mathbf{A}_i$, indicating whether each parameter satisfies its non-exploitable condition $\mathbf{U}$. Each $s_{i,k}$ represents whether $a_{i,k}$ satisfies its corresponding non-exploitable condition $\mathbf{U}$. \\ \hline

$\boldsymbol{\Phi}_{tr}$ & Transfer Constraints & For each pair of adjacent methods $(m_i, m_{i+1})$ on $\mathbf{P}$, $\boldsymbol{\Phi}_{tr}^i$ denotes the constraint on the parameters of $m_i$ that must be satisfied for the execution to proceed to $m_{i+1}$. $\boldsymbol{\Phi}_{tr}$ denotes the collection of all such constraints, ensuring execution from source $m_1$ to sink $m_n$. \\ \hline

$\boldsymbol{\Phi}_{tg}$ & Trigger Constraints & $\boldsymbol{\Phi}_{tg}$ specifies that, for the sink method $m_n$ on $\mathbf{P}$, its parameter states $\mathbf{S}_n$ must satisfy the conditions required to trigger the vulnerability. \\ \hline
\end{tabularx}
\vspace{0.4pt}
\caption{Core concepts for vulnerability detection.}
\label{tab:definitions}
\end{table*}

\subsection{LLM-Based Approaches}
\label{challenges_llms}

LLMs, with their powerful code understanding, logical reasoning, and planning capabilities, have been increasingly applied to the field of security analysis~\cite{liu2024vuldetectbench,mao2025towards,ma2025psyscam}. 
Recent studies have explored their potential in tasks such as zero-shot vulnerability identification, context-aware reasoning, and explainable detection. 
For instance, VulDetectBench \cite{liu2024vuldetectbench} designed a five-stage benchmark to evaluate LLMs’ ability to identify, classify, and localize vulnerabilities, showing that while LLMs achieve over 80\% accuracy in simple classification tasks, their performance drops below 30\% in fine-grained vulnerability localization. 
Similarly, LLMVulExp \cite{mao2025towards} combined Chain-of-Thought (CoT) prompting with LoRA fine-tuning to improve explainability, achieving over 90\% F1-score on the SeVC dataset. 
Although LLMs demonstrate significant potential in security analysis, their effectiveness in vulnerability detection faces several critical limitations.

LLMs are constrained by context limitations. Their fixed context windows cannot accommodate large codebases, and even within these limits, lengthy inputs cause substantial accuracy degradation~\cite{kaniewski2025systematic}, making it difficult to reliably capture complex call relationships across repositories.

Agent-based architectures, where LLMs autonomously drive each analysis step, suffer from severe instability. The excessive freedom granted to LLMs makes system behavior unpredictable, even to developers, resulting in high false-positive rates as models arbitrarily shift focus or lose critical context~\cite{wang2024llmdfa,zibaeirad2025reasoning}.

SAST-LLM hybrid approaches often fail to leverage LLMs' full potential. SAST-centric solutions use LLMs merely to supplement rules, inheriting SAST's limitations while underutilizing semantic reasoning capabilities. Conversely, LLM-centric solutions typically adopt agent architectures, reintroducing the instability issues described above.

These key limitations seriously hinder the practical deployment of LLMs in industrial-grade vulnerability detection, making it difficult for them to provide practical value in real production environments.

\subsection{Problem Statement}
\label{problem_statement}

We cast vulnerability detection as a constraint-solving problem: for any call chain leading to a dangerous sink method, a vulnerability is deemed to exist if all associated constraints are satisfied; otherwise, no vulnerability is present.
To better characterize the vulnerability detection process, Table~\ref{tab:definitions} summarizes the core concepts and their corresponding notations.

\subsubsection{Formal Problem Definition}
\label{formal_definition}

Based on the above definitions, we formalize the vulnerability detection problem as a constraint satisfaction problem:




Transfer constraints require that the input values guarantee the complete execution of the main path $\mathbf{P}$ from the source method $m_1$ to the sink method $m_n$, without premature termination. Trigger constraints require that, once the sink method $m_n$ is reached, the states of its critical parameters $\mathbf{S}_n$ satisfy the conditions necessary to activate the vulnerability.

Taking the running example in Listing~\ref{runing_example_lst} as an illustration, the transfer constraints $\boldsymbol{\Phi}_{tr}$ require that the user input $\mathit{Input}$ drives execution along the main path $\mathbf{P}$ from the source $m_1 = \textit{doGet}$ to the sink method $m_n = \textit{readString}$ without premature termination. The trigger constraints $\boldsymbol{\Phi}_{tg}$ require that, when $m_n$ ($\textit{readFile}$) executes, $\mathbf{S}_n$, the state of its critical parameters $\mathbf{A}_n$ (whose corresponding argument in the code is $\textit{Paths.get(path)}$), satisfies the exploitation condition of the sink method. Here, the exploitation condition of the sink method is that $\textit{Paths.get(path)}$ does not satisfy the corresponding $\mathbf{U}$ (i.e., ".." in its value has been safely filtered).

\begin{tcolorbox}[title=Vulnerability Detection Constraint,
                  label={prob:vulnerability_detection},
                  colback=gray!5,
                  colframe=black!60,
                  fonttitle=\bfseries,
                  breakable,
                  enhanced]

\textbf{Input}: User-provided values along the call chain. \\
\textbf{Constraints}: $\boldsymbol{\Phi}_{tr} \;\wedge\; \boldsymbol{\Phi}_{tg}$ \\
\textbf{Problem Statement}:
Find whether there exists an input assignment $\mathit{Input}$ that satisfies:
\[
\mathit{Input} \models \boldsymbol{\Phi}_{tr} \;\wedge\; \boldsymbol{\Phi}_{tg}.
\]

\end{tcolorbox}

Whether the trigger constraints $\boldsymbol{\Phi}_{tg}$ are satisfied depends on the state $\mathbf{S}_n$ of the critical types at the sink method. When there are multiple critical parameters in $\mathbf{A}_n$, $\boldsymbol{\Phi}_{tg}$ may be satisfied through a logical combination of their states, rather than requiring all parameters to uniformly satisfy or violate their non-exploitable conditions $\mathbf{U}$. For example, in a file reading method with multiple path parameters that are concatenated, $\boldsymbol{\Phi}_{tg}$ is satisfied if any one of them violates its corresponding $\mathbf{U}$.

\subsubsection{Solving Algorithm Framework}
\label{algorithm_framework}

Vulnerability detection constraint solving requires solving both the transfer constraints $\boldsymbol{\Phi}_{tr}$ and the trigger constraints $\boldsymbol{\Phi}_{tg}$.
The transfer constraints $\boldsymbol{\Phi}_{tr}$ hold if, for every pair of adjacent methods $(m_i, m_{i+1})$ on the main path $\mathbf{P}$, the corresponding constraint $\boldsymbol{\Phi}_{tr}^i$ is satisfied. Formally,
\[
\boldsymbol{\Phi}_{tr} = \{\boldsymbol{\Phi}_{tr}^i \mid i = 1,2,\ldots,n-1\}.
\]

As for the trigger constraints, $\boldsymbol{\Phi}_{tg}$ stipulates that at the sink method $m_n$, the parameter states $\mathbf{S}_n$ must satisfy the conditions that render the sink exploitable.

Inspired by the human expert auditing process, instead of solving this complex problem directly, we decompose it into a sequence of subtasks. Each subtask focuses on an adjacent method pair $(m_i, m_{i+1})$, with the goal of ensuring that the transfer constraint $\boldsymbol{\Phi}_{tr}^i$ is satisfied while simultaneously deriving the parameter state $\mathbf{S}_{i+1}$ of the callee $m_{i+1}$. The result of each subtask then serves as the input condition for the next one, enabling the derivation of $\mathbf{S}_{i+2}$, and so on, until $\mathbf{S}_n$ is obtained. Finally, $\mathbf{S}n$ is used to check whether the trigger constraints $\boldsymbol{\Phi}_{tg}$ are satisfied, thereby determining the existence of a vulnerability.

\begin{tcolorbox}[title=Subtask,
                  label={prob:subtask_solving},
                  colback=gray!5,
                  colframe=black!60,
                  fonttitle=\bfseries,
                  breakable,
                  enhanced]

\textbf{For:} adjacent method pair $(m_i, m_{i+1})$ \\
\textbf{Given:} 
Parameter state $\mathbf{S}_i$ of $m_i$ (whether $a_{i,k}\in\mathbf{A}_i$ satisfies $\mathbf{U}$); Source code of method $m_i$; Source code of $m_i$'s branch methods $\mathbf{N}_i$; Source code of $m_i$'s branch trees $\mathbf{T}_i$ \\
\textbf{Objective:} 
Derive parameter state $\mathbf{S}_{i+1}$ of $m_{i+1}$ (whether $a_{i+1,k}\in\mathbf{A}_{i+1}$ satisfies $\mathbf{U}$) under the condition that $\boldsymbol{\Phi}_{tr}^i$ is satisfied.

\end{tcolorbox}

It is important to note that if every subtask corresponding to adjacent methods $(m_i, m_{i+1})$ is successfully solved, then both the transfer and trigger constraints are solved, allowing us to determine whether a vulnerability exists.

Based on the human expert auditing process, we identify that solving the subtask problem effectively requires the following three key capabilities:
\textit{(a) Branch Method Analysis}: This task analyzes the branch trees $\mathbf{T}_i$ to extract the semantics of $m_i$'s branch methods $\mathbf{N}_i$, transforming the large amount of complex code on the branch trees into a simple conclusions of what $m_i$ does, in order to determine whether these branch methods affect $\mathbf{S}_{i+1}$.
\textit{(b) Context Maintenance}: This task leverages the result $\mathbf{S}_i$ from the previous subtask together with the outcome of Branch Method Analysis, further distilling them into precise contextual information necessary for accurately deriving $\mathbf{S}_{i+1}$, eliminating the need to revisit the extensive and complex code from previous analysis.
\textit{(c) Main Path Analysis}: Based on the context obtained from Context Maintenance and the source code of $m_i$, this task ultimately derives $\mathbf{S}_{i+1}$.

Through systematic design, \sname replaces unstable LLM agents with a deterministic framework mirroring human auditing to achieve these capabilities.
The details are presented in Section~\ref{SBC-Solving}.

\lstdefinestyle{javastyle}{
    language=Java,
    basicstyle=\ttfamily\footnotesize,
    keywordstyle=\color{blue}\bfseries,
    commentstyle=\color{gray}\itshape,
    stringstyle=\color{red},
    numberstyle=\tiny\color{gray},
    numbers=left,
    numbersep=3pt,
    frame=none,
    backgroundcolor=,
    breaklines=true,
    breakatwhitespace=true,
    tabsize=2,
    showstringspaces=false,
    captionpos=b,
    aboveskip=0.5em,
    belowskip=0.5em,
    xleftmargin=0pt,
    xrightmargin=0pt,
    framexleftmargin=0pt,
    framexrightmargin=0pt,
    lineskip=0pt
}

\lstdefinelanguage{PseudoJSON}{
  alsoletter=-,
  morestring=[b]",
  morecomment=[l]{//},
  morekeywords={true,false,null},
}

\lstdefinestyle{jsonstyle}{
    language=PseudoJSON,
    basicstyle=\ttfamily\footnotesize\color[rgb]{0.2,0.2,0.2},    
    keywordstyle=\color[rgb]{0.09,0.28,0.48}\bfseries,             
    commentstyle=\color[rgb]{0.42,0.42,0.42}\itshape,              
    stringstyle=\color[rgb]{0.62,0.13,0.22},                       
    numberstyle=\tiny\color[rgb]{0.5,0.5,0.5},                     
    emph={methods,className,def,code,args,branchs,
          snippetOfCalled,invokerOfCalled,memberVariables,
          passRelationShip,pollutedPosition},                      
    emphstyle=\color[rgb]{0.11,0.49,0.35}\bfseries,                
    backgroundcolor=\color{white},
    frame=none,
    framerule=0.8pt,
    rulecolor=\color[rgb]{0.82,0.82,0.82},                         
    numbers=left,
    numbersep=5pt,
    breaklines=true,
    breakatwhitespace=true,
    tabsize=4,
    showstringspaces=false,
    xleftmargin=12pt,
    aboveskip=8pt,
    belowskip=4pt
}

\section{Design}
\label{design}

This section presents the system design of \sname, a language-independent framework for vulnerability detection through constraint solving. We first provide an overview of the framework architecture and its key modules, then detail the design of each module. The design realizes the three key capabilities identified in Section~\ref{algorithm_framework}, enabling effective vulnerability detection across diverse programming languages.




\subsection{Overview}
\label{overview}

\sname introduces a semantic-based constraint-solving framework that guides LLMs to perform vulnerability detection in a predictable and controllable manner, closely resembling the reasoning process of human experts.
Unlike traditional architectures --- which often grant LLMs excessive autonomy and introduce instability in complex audits due to randomness \cite{khare2025understanding,liu2024vuldetectbench,zibaeirad2025reasoning,steenhoek2024toerrismachine} --- our approach systematically models the workflow of security experts.
It decomposes vulnerability discovery into a sequence of subtasks with dedicated solving mechanisms, constraining LLM analysis within a well-defined semantic space.
In doing so, the framework preserves the semantic understanding capabilities of LLMs while mitigating the uncertainty of free-form reasoning through formalized constraints, ultimately combining human-level precision with automated scalability in vulnerability detection.

\begin{figure*}
    \centering
    \includegraphics[width=\textwidth]{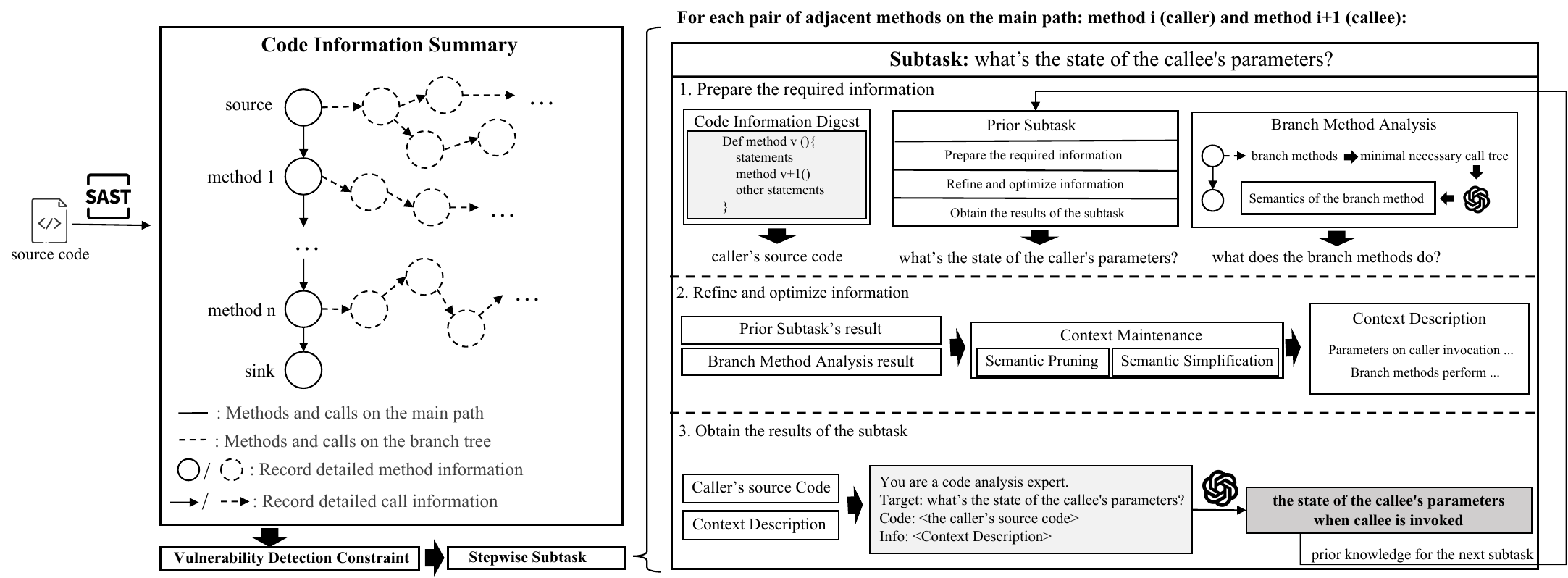} 
    \caption{The overall workflow of \sname}
    \label{fig:Architecture Diagram of sname}
\end{figure*}

Figure~\ref{fig:Architecture Diagram of sname} presents the overall workflow of \sname, consisting of two components: 

\noindent\textbf{Code Information Summary Generation}. \sname employs SAST to preprocess source code into an intermediate representation capturing potential vulnerability call chains and metadata (Section~\ref{CIS_G}). This provides a universal input format enabling the same constraint-solving logic across different languages and vulnerability types.

\noindent\textbf{Subtask Resolution Module}. This core component decomposes constraint solving into subtasks, each analyzing an adjacent caller–callee pair along the main path (Section~\ref{algorithm_framework}). Each subtask infers the callee's parameter state using the caller's state from the preceding subtask. By iteratively propagating states along the path, the framework derives the sink's parameter state and evaluates Trigger Constraints.

Each subtask relies on three capabilities: Branch Method Analysis extracts branch method semantics; Context Maintenance integrates previous results into a semantic abstraction of the caller's state; Main Path Analysis uses this context to infer the callee's parameter state.

\subsection{Code Information Summary Generation}
\label{CIS_G}

\sname employs SAST to generate code information summaries that serve as standardized inputs for constraint solving. This decouples raw code from the solving logic, allowing the same logic to operate across different languages and vulnerability types by working on uniform summaries rather than handling language-specific syntax.

The summary contains two key elements: call chains and their metadata.

\vspace{0.1cm}
\noindent\textbf{Call Chains}. The summary can include any call chain reaching a dangerous sink method, even those SAST deems unexploitable. This is especially important when SAST has high false negative rates. By treating SAST solely as an information provider rather than a decision maker, this approach avoids the inherent limitations of SAST-based vulnerability discovery. \sname outputs exploitability judgments with detailed reasoning for each chain.



\vspace{0.1cm}
\noindent\textbf{Metadata}. Each call chain requires the following metadata: method details including names, source code, signatures (static/constructor), parameter types and names for main methods and branch methods; caller-callee relationships between main methods and branch methods; detailed types of each method's parameters used to identify critical types for constraint solving, where encapsulated types must also be recorded for parameters whose types encapsulate other types; and data flow propagation between methods and within branch methods, which, although SAST cannot capture high-level semantics, assist subsequent analysis steps.

Listing~\ref{lst:schema-summary} presents a representative schema for summarizing code information.
It aggregates function-level metadata --- such as \textit{className}, call relationships, and variables --- providing the semantic context needed for large language models (LLMs) to analyze behavior and surface potential vulnerability indicators.

\subsection{Semantic-Based Constraint Solving}
\label{SBC-Solving}

As described in Section~\ref{algorithm_framework}, constraint solving reduces to a sequence of subtasks, each requiring support from three complementary modules. Branch Method Analysis captures the semantics of branch methods associated with the caller, Context Maintenance integrates prior task outcomes with branch information to maintain an accurate analysis context, and Main Path Analysis utilizes this context to infer the callee’s parameter state. The following subsections detail the implementation of these modules.

\subsubsection{Branch Method Analysis}
\label{branch_analyze}

\begin{lstlisting}[style=jsonstyle,caption={A typical schema of Code Information Summary},label={lst:schema-summary}, float=t]
{
  "methods": [
    {
      "className": "<Class name the method belongs to>",
      "def": "<Method definition>",
      "code": "<Method source code>",
      "args": [
        {
          "name": "<Method parameter name>",
          "type": "<Method parameter type>"
        }
        // ...(other parameters)
      ],
      "branchs": [
        "<Information of branch methods>"
      ],
      "snippetOfCalled": "<Expression of the callee in the caller's code>",
      "invokerOfCalled": "<Expression of the callee instance in the caller's code>",
      "memberVariables": [
        {
          "name": "<Member variable name of the callee>",
          "type": "<Member variable type of the callee>"
        }
        // ...(other member variables)
      ],
      "passRelationShip": "<Mapping between actual arguments and formal parameters of the callee>",
      "pollutedPosition": "<Taint propagation relationship>"
    }
    // ...(subsequent main methods)
  ]
}
\end{lstlisting}

\begin{figure*}
    \centering
    \includegraphics[width=0.99\linewidth]{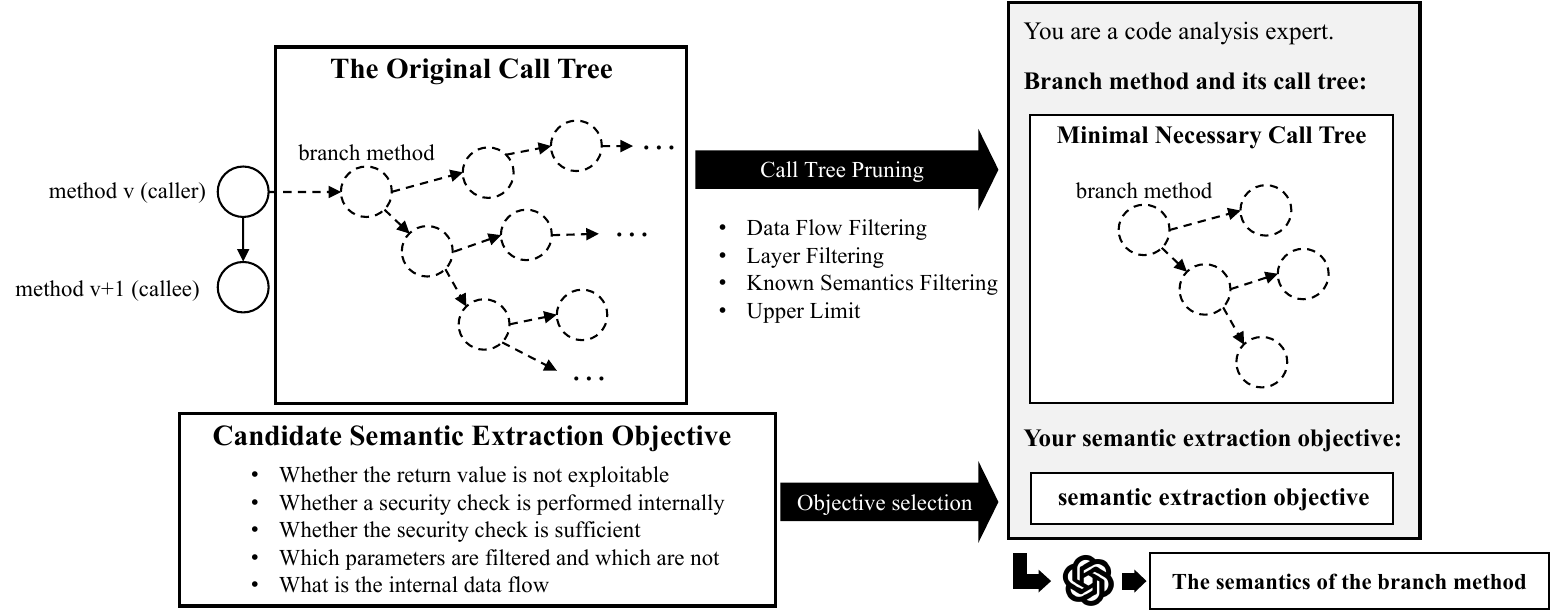} 
    \caption{The workflow of Branch Method Analysis}
    \label{fig:Branch Method Analyze}
\end{figure*}

Branch Method Analysis focuses on extracting the semantics of branch methods, providing supplementary information that is refined by the Context Maintenance module and later consumed by Main Path Analysis. Its purpose is to condense extensive branch trees into concise conclusions about what each branch method does, allowing the LLM to focus on the main method without analyzing complex branch code when subsequently analyzing the main method.

Its internal workflow is illustrated in Figure~\ref{fig:Branch Method Analyze}.
This module comprises two components, namely Call Tree Pruning and Objective Selection, which respectively address the two core challenges of Branch Method Analysis. Specifically, the first challenge concerns how to minimize the amount of code the LLM must process to improve accuracy, while the second addresses what information should be extracted for main method analysis without causing information loss.
Call Tree Pruning reduces the branch tree to the minimal call structure required for semantic extraction, while Objective Selection identifies the specific semantic aspects to be extracted for use in Main Path Analysis.

Together, these steps ensure that only the essential code is analyzed and that the extracted semantics are directly aligned with the needs of subsequent analysis.

\vspace{0.1cm}
\noindent\textbf{Call Tree Pruning}
As defined in Table~\ref{tab:definitions}, each branch method transitively invokes other methods, forming a call tree referred to as the branch tree. To extract the semantics of a branch method, we prune this tree to retain only the minimal set of necessary methods for analysis, thereby enabling precise semantic extraction.
Specifically, \sname applies the following filtering strategies to obtain the minimal necessary call tree.

\emph{Data Flow Filtering}. Branch Method Analysis focuses only on methods relevant to the data flow of the analysis target. Methods in the call tree without data-flow connections to the target are excluded. For example, when analyzing whether the return value of a branch method is filtered, any method unrelated to the data flow leading to that return value is discarded.

\emph{Layer Filtering}. Deeper layers in the call tree are semantically farther from the current branch method. \sname uses breadth-first traversal to prioritize shallower layers, discarding code in excessively deep layers.

\emph{Known Semantics Filtering}. Methods with widely recognized semantics (e.g., built-in or common framework methods) are already learned by the LLM during pretraining. Their source code is thus unnecessary and excluded from analysis.

\emph{Upper Limit}. To prevent performance degradation from excessive code volume, \sname enforces an upper threshold on the number of retained methods.


\vspace{0.1cm}
\noindent\textbf{Objective Selection}
Like human security experts who focus on different aspects depending on the method type, \sname determines the critical semantics to extract from a branch method based on its signature—specifically, its input and output types.
The branch method types and their corresponding analysis objectives are as follows:

\emph{Critical Type Constructors / Methods Returning Critical Types}. For methods that construct or return critical types, the analysis identifies which input parameters contribute to the returned critical type using data-flow information from the code information summary. Each contributing parameter is then independently examined via LLM to determine whether its assignment path satisfies non-exploitability conditions.
For example, if the return value derives from two parameters, the analysis may conclude that one parameter's assignment path is filtered while the other is not. The final result precisely identifies which parameters reach the return value without filtering.

\emph{Constructors of Encapsulated Types / Methods Returning Encapsulated Types}. For methods that construct or return types encapsulating critical types, the analysis objective remains the same as above, but focuses on the critical types encapsulated within the constructed or returned object.

\emph{Methods with Critical-Type Parameters}. For methods whose parameters include critical types, the analysis identifies which input parameters contribute to each critical-type parameter using data-flow information. Each contributing parameter is independently examined to determine whether its assignment path satisfies non-exploitability conditions.
For example, if a critical-type parameter receives data from two other parameters, the analysis may determine that one parameter's path is filtered while the other is not. The result identifies which parameters reach each critical-type parameter without filtering.

\emph{Methods with Encapsulated-Type Parameters}. For methods whose parameters include types that encapsulate critical types, the analysis objective remains the same as above, but focuses on the critical types encapsulated within these parameters.

\emph{Methods Returning Boolean}. For methods returning boolean values, the analysis determines whether the logic governing the return value reveals the relationship between critical types and non-exploitability conditions. This relationship may be indirect rather than explicit. For example, a method may directly check whether a parameter is in a specific whitelist, which indirectly determines whether the critical type satisfies non-exploitability conditions.

\emph{Other Branch Methods}. For branch methods not matching the above categories, the analysis directly derives their internal data-flow propagation relationships from the code information summary.

Through Call Tree Pruning, the minimal necessary code required for analysis is obtained, and through Objective Selection, the specific semantics to be extracted are identified. These processes are performed entirely through \sname's deterministic code logic in a precise and efficient manner, without requiring LLM involvement. Based on this systematically prepared information, \sname constructs prompts and invokes the LLM to perform the final step of Branch Method Analysis, producing semantic summaries for each required branch method.

\subsubsection{Context Maintenance}
\label{context_maintenance}

Context Maintenance provides all contextual information required by Main Path Analysis beyond the code itself. It manages all prior analysis results, including conclusions from earlier Main Path Analysis and Branch Method Analysis, and distills them into precise and complete contextual summaries. This enables the LLM to perform Main Path Analysis by focusing solely on the caller's code in each caller-callee pair, without examining any other code, thereby achieving more accurate analysis.

\begin{figure*}
    \centering
    \includegraphics[width=\textwidth]{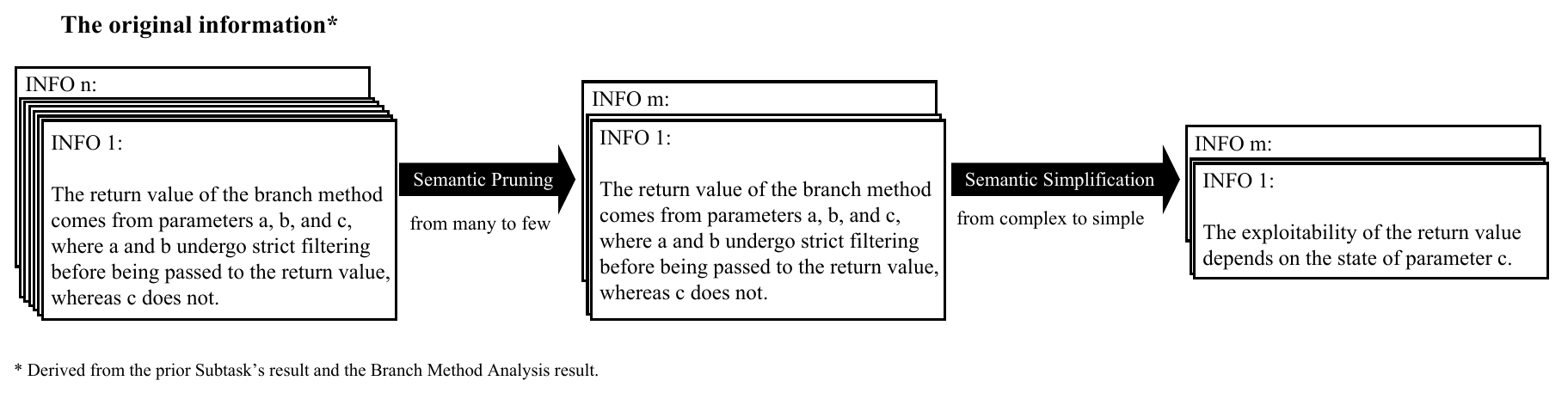} 
    \caption{The workflow of Context Maintenance}
    \label{fig:Context Maintenance}
\end{figure*}



Specifically, Context Maintenance achieves this by applying semantic pruning, which reduces the quantity of contextual information by removing irrelevant or redundant elements to retain only what is necessary, and semantic simplification, which reduces the complexity of content by condensing elaborate semantic descriptions into shorter, more comprehensible forms. This process is illustrated in Figure~\ref{fig:Context Maintenance}.




\noindent\textbf{Semantic Pruning Strategies}
Semantic pruning aims to reduce the volume of contextual information by systematically eliminating elements that do not contribute to the current analysis. This is achieved by evaluating the relevance of parameters and branch methods based on their actual usage in subsequent analysis.

\emph{Parameter Pruning.} Not all parameters from the caller require inclusion in the context for analyzing the callee's parameter state. Thanks to \sname's architectural design, all code required for subsequent Main Path Analysis and Branch Method Analysis can be determined in advance, including all main methods and branch methods. Parameters that do not appear in any of this code can be safely excluded. For instance, consider a scenario where prior analysis concludes that \texttt{x}, a critical-type member encapsulated within a caller's parameter, satisfies non-exploitability conditions. If \texttt{x} is never referenced in the code to be analyzed subsequently (neither in main methods nor in branch methods), its state does not need to be preserved in the context, as it has no bearing on the analysis of the callee.

\emph{Branch Method Pruning.} Not all branch methods invoked by the caller are necessary for analyzing the callee's parameter state. The relevance of a branch method is determined by whether it affects the data flow from the caller's parameters to the callee's parameters. Branch methods that have no data-flow connection to this propagation path can be excluded, as their semantics do not influence the analysis outcome. Additionally, branch methods whose behavior is encoded in the LLM's prior knowledge (such as standard library methods or widely-used framework APIs) need not be included in the context, since the model already possesses sufficient understanding of their semantics without explicit code provision. These filtering criteria mirror the data flow filtering and known semantics filtering strategies employed in branch tree reduction during Branch Method Analysis.

\emph{Semantic Simplification Strategies}
Semantic simplification transforms complex code semantics into more concise and comprehensible representations, enabling the LLM to process contextual information more efficiently. This is primarily applied to the semantics of branch methods, where detailed implementation logic is abstracted into higher-level semantic descriptions that preserve analytical relevance while reducing cognitive load.

\emph{Propagation Simplification.} Within branch methods, internal data-flow relationships that do not affect whether critical types satisfy non-exploitability conditions are abstracted as direct assignments. The intermediate operations performed during propagation are omitted from the semantic description, as they do not alter the exploitability state of the data. This simplification allows the LLM to focus on the essential data-flow connections without being distracted by implementation details that are irrelevant to vulnerability analysis.

\emph{Filtering Simplification.} Any method logic that ensures critical types satisfy non-exploitability conditions is uniformly represented as a security check, regardless of whether the filtering effect is achieved through explicit sanitization operations or incidentally through other operations. This abstraction provides a consistent semantic representation that clearly indicates the presence of security-relevant transformations, enabling the LLM to recognize protective measures without needing to interpret the detailed mechanisms by which they are implemented.

\emph{Exploitability Judgment Simplification.} When a critical type is derived from multiple sources with varying exploitability states (some filtered and some unfiltered), its semantic description focuses exclusively on the unfiltered sources. Filtered sources are omitted from the description, as their non-exploitable state means they do not contribute to potential vulnerabilities. This simplification directs the LLM's attention to the sources that actually pose security risks, streamlining the analysis by eliminating information about data flows that have already been deemed safe.

Through semantic pruning and semantic simplification, the context becomes both easier for the LLM to understand and more instructive for analysis. The simplified semantics not only reduce complexity but also act as step-by-step guidance: if the LLM needs to judge whether a critical type satisfies non-exploitability conditions, the context explicitly tells it which variables must be analyzed first. When those variables in turn depend on others, the context again provides clear instructions, forming a recursive chain of guidance. This process greatly improves the accuracy and stability of LLM analysis.

\subsubsection{Main Path Analysis}



The purpose of Main Path Analysis is to determine the parameter state of the callee in each caller-callee invocation pair. Specifically, the callee's parameter state is represented by whether its critical parameters satisfy their corresponding Non-exploitable Conditions. 

Main Path Analysis is an incremental process: the analyzed state of each callee is passed to Context Maintenance, which transforms it into known conditions for the next invocation pair, where the previous callee becomes the new caller. When the callee is a sink method, the result of Main Path Analysis reveals the state of the sink's critical parameters, thereby determining whether the sink can be exploited, i.e., whether the trigger constraints are satisfied.

In Main Path Analysis, \sname provides only the caller's source code directly. All other necessary information is supplemented through systematically constructed textual descriptions from Context Maintenance, including the state of the caller's parameters (from preceding Main Path Analysis) and the semantics of the caller's branch methods (from Branch Method Analysis).

This design greatly reduces the amount of code that must be supplied. The LLM only needs to focus on the caller’s code, while the complex logic of branch methods and preceding main methods has already been analyzed earlier and distilled into concise conclusions provided as context. As a result, the LLM can rely on these contextual conclusions rather than re-analyzing deeper code, making its task simpler and its analysis far more accurate.

The final Main Path Analysis yields the parameter state of the sink method at its invocation point, which directly determines whether the trigger constraints are satisfied and thus whether a vulnerability exists. Trigger constraints are defined as logical combinations of the sink's parameter states. For example, if a file-reading method concatenates two path fragment parameters into the actual file path, the trigger constraints are satisfied when at least one parameter fails to meet its non-exploitability condition. Once the sink's parameter state is derived, verifying the trigger constraints becomes straightforward.

\lstdefinestyle{javastyle}{
    language=Java,
    basicstyle=\ttfamily\footnotesize,
    keywordstyle=\color{blue}\bfseries,
    commentstyle=\color{gray}\itshape,
    stringstyle=\color{red},
    numberstyle=\tiny\color{gray},
    numbers=left,
    numbersep=3pt,
    frame=none,
    backgroundcolor=,
    breaklines=true,
    breakatwhitespace=true,
    tabsize=2,
    showstringspaces=false,
    captionpos=b,
    aboveskip=0.5em,
    belowskip=0.5em,
    xleftmargin=0pt,
    xrightmargin=0pt,
    framexleftmargin=0pt,
    framexrightmargin=0pt,
    lineskip=0pt
}

\section{Evaluation}
\subsection{Experimental Design}
To comprehensively evaluate the effectiveness and practicality 
of \sname, our constraint solving framework for vulnerability 
detection, we designed a series of systematic experiments. These experiments aim to answer the following key research questions:

\begin{itemize}[leftmargin=0pt, topsep=2pt, itemsep=0pt, label=]
\item \textbf{RQ1}: Compared to existing SAST tools and LLM-based vulnerability detection methods, does our approach demonstrate superior performance in core metrics such as accuracy, precision, recall, and F1-score?
\item \textbf{RQ2}: What is the contribution of core modules such as Branch Method Analysis, Context Maintenance, and Main Path Analysis to overall performance? Which components are key factors for performance improvement?
\item \textbf{RQ3}: Can our approach effectively discover actual security vulnerabilities in real large-scale open-source projects? Does it meet the requirements for practical deployment?
\end{itemize}

\subsection{Dataset}
The experimental evaluation employs a dual-source data collection strategy to ensure comprehensive assessment: (1) the OWASP Benchmark serves as the primary standardized dataset for comprehensive evaluation, and its widespread adoption enables direct performance comparison with existing approaches; (2) a curated selection of high-profile open-source repositories provides validation in authentic deployment scenarios. This methodological approach facilitates rigorous comparative analysis within standardized benchmarking frameworks while simultaneously demonstrating practical applicability in real-world environments. From the OWASP Benchmark, we extracted 1,023 labeled test cases covering common high-risk vulnerability types, including SQL Injection, Command Injection, and Path Traversal vulnerabilities.

\subsubsection{Experimental Environment and Configuration}

To ensure reproducibility and fairness of experiments, we standardized the experimental environment configuration. For LLM configuration, the evaluation primarily employs GPT-4, GPT-4-Turbo, DeepSeek-V3(250324) and Kimi-K2(250905) models, with temperature parameter set to 0.3. For SAST tool configuration, Tabby serves as the underlying static analysis component~\cite{chen2023tabby}, and the results produced by Tabby are further transformed into Code Information Summaries that conform to the format specified in Section~\ref{CIS_G}.

\section{Experimental Results and Analysis}
In this section, we will sequentially answer the research questions posed earlier and provide detailed analysis of the experimental results. We evaluate our method using standard classification metrics including Accuracy, Precision, Recall, and F1-score, along with vulnerability type-specific performance analysis to comprehensively assess detection capabilities across different vulnerability categories.

\subsection{RQ1: Effectiveness of Current Method}
In this section, we evaluated the effectiveness of our proposed method and compared it with existing baseline methods. The experimental results demonstrate that our method performs excellently across all core metrics, particularly achieving a low false positive rate while maintaining high recall.

\subsubsection{Performance Evaluation and Analysis}

Table~\ref{tab:overall-performance} presents the overall performance of \sname on the test set. The method achieves 97.85\% accuracy and 97.97\% F1-score, with a particularly notable characteristic: 100\% recall with 96.02\% precision. Out of 1023 test samples, \sname correctly identifies all 531 vulnerable samples and 470 out of 492 benign samples, resulting in only 22 false positives and zero false negatives. This demonstrates that \sname ensures zero missed vulnerabilities while maintaining high precision.

\begin{table}[H]
\renewcommand{\arraystretch}{1.2}
\setlength{\tabcolsep}{15pt}
  \caption{Performance of \sname on Overall Test Set}
  \label{tab:overall-performance}
  \centering
    \begin{tabular}{cccc}
    \hline
     \textbf{Accuracy} & \textbf{Precision} & \textbf{Recall} & \textbf{F1} \\
    \hline
    97.85\% & 96.02\% & 100\% & 97.97\% \\
    \hline
    \noalign{\vskip 3pt}
    \hline
    \textbf{TP} & \textbf{FP} & \textbf{TN} & \textbf{FN} \\
    \hline
    531 & 22 & 470 & 0 \\
    \hline
  \end{tabular}
\end{table}

To understand how \sname performs across different attack vectors, Table~\ref{tab:vulnerability-types} breaks down performance by vulnerability category. A consistent pattern emerges: the method maintains 100\% recall across all types, confirming reliable detection capability regardless of vulnerability nature. The F1-scores further validate this effectiveness, with command injection reaching 98.44\% and SQL injection achieving 98.37\%. Even for path traversal—the most challenging type where file system operations with variables frequently appear in both vulnerable and legitimate code—the F1-score remains at 96.73\%, demonstrating robust detection performance across diverse vulnerability patterns.

\begin{table}[H]
\renewcommand{\arraystretch}{1.25}
  \caption{Performance of Our Method on Different Vulnerability Types}
  \label{tab:vulnerability-types}
  \begin{adjustbox}{width=0.97\linewidth, center}
    \begin{tabular}{lcccc}
    \hline
    \textbf{Vulnerability Type} & \textbf{Acc.} & \textbf{Prec.} & \textbf{Rec.} & \textbf{F1} \\
    \hline
    Command Injection & 98.41\% & 96.92\% & 100\% & 98.44\% \\
    Path Traversal    & 96.64\% & 93.66\% & 100\% & 96.73\% \\
    SQL Injection     & 98.21\% & 96.80\% & 100\% & 98.37\% \\
    \hline
  \end{tabular}
  \end{adjustbox}
\end{table}

These results validate the effectiveness of \sname in real-world deployment scenarios. The zero false negative rate ensures comprehensive security coverage, while the consistently high precision across all vulnerability types (above 93.66\%) means security teams can confidently act on detected vulnerabilities without being overwhelmed by false alarms. The combination of complete vulnerability coverage and practical precision establishes \sname as an effective and reliable vulnerability detection approach.

\begin{table*}[t]
\renewcommand{\arraystretch}{1.1}
  \centering
  \caption{Performance Comparison of Different Methods on Various Vulnerability Types}
  \label{tab:comprehensive-comparison}
  \begin{adjustbox}{width=0.9\linewidth, center}
  \begin{tabular}{llcccc}
    \hline
    \textbf{Vulnerability Type} & \textbf{Method} & \textbf{Acc.} & \textbf{Prec.} & \textbf{Rec.} & \textbf{F1} \\
    \hline
    {\textbf{Command Injection}} 
    & \textbf{\sname (Kimi-K2)} & \textbf{98.41\%} & \textbf{96.92\%} & \textbf{100\%} & \textbf{98.44\%} \\
    & \sname (DeepSeek-V3) & 93.63\% & 88.73\% & 100\% & 94.03\% \\
    & \sname (GPT-4o) & 90.44\% & 84.00\% & 100\% & 91.30\% \\
    & \sname (GPT-4-turbo) & 92.83\% & 87.50\% & 100\% & 93.33\% \\
    & CodeQL & 56.00\% & 53.00\% & 77.00\% & 63.00\% \\
    & CWE-DF & 48.00\% & 48.00\% & 100\% & 65.00\% \\
    & Best-Rated Agent & 74.30\% & -- & -- & -- \\
    \hline
    {\textbf{Path Traversal}} 
    & \sname (Kimi-K2) & 96.64\% & 93.66\% & 100\% & 96.73\% \\
    & \textbf{\sname (DeepSeek-V3)} & \textbf{98.88\%} & \textbf{97.79\%} & \textbf{100\%} & \textbf{98.88\%} \\
    & \sname (GPT-4o) & 92.54\% & 86.93\% & 100\% & 93.01\% \\
    & \sname (GPT-4-turbo) & 94.03\% & 89.26\% & 100\% & 94.33\% \\
    & CodeQL & 52.00\% & 50.00\% & 100\% & 67.00\% \\
    & CWE-DF & 48.00\% & 48.00\% & 100\% & 64.00\% \\
    & Best-Rated Agent & 70.50\% & -- & -- & -- \\
    \hline
    {\textbf{SQL Injection}} 
    & \textbf{\sname (Kimi-K2)} & \textbf{98.21\%} & \textbf{96.80\%} & \textbf{100\%} & \textbf{98.37\%} \\
    & \sname (DeepSeek-V3) & 96.23\% & 93.47\% & 100\% & 96.63\% \\
    & \sname(GPT-4o) & 92.86\% & 88.31\% & 100\% & 93.79\% \\
    & \sname(GPT-4-turbo) & 93.25\% & 88.89\% & 100\% & 94.12\% \\
    & CodeQL & 57.00\% & 54.00\% & 100\% & 70.00\% \\
    & CWE-DF & 52.00\% & 52.00\% & 100\% & 68.00\% \\
    & Best-Rated Agent & 67.80\% & -- & -- & -- \\
    \hline
  \end{tabular}
  \end{adjustbox}
  
  \vspace{0.2cm}
  \raggedright
  \small
  Note: "--" indicates that the corresponding metric was not reported in the original study.
\end{table*}

\subsubsection{Comparative Analysis with Baseline Methods}

To comprehensively evaluate the effectiveness of \sname, we compare its performance against established baselines from prior work~\cite{khare2025understanding, shashwat2024preliminary}. The comparison includes three types of approaches: (1) CodeQL, a widely-adopted industrial static analysis tool representing traditional SAST methods; (2) CWE-DF, the best-performing configuration in their experiments across different LLM models and prompting strategies; and (3) Best-Rated Agent, the best-performing AI agent in their experimental evaluation. These baselines represent recent advances in both conventional and LLM-based vulnerability detection approaches.

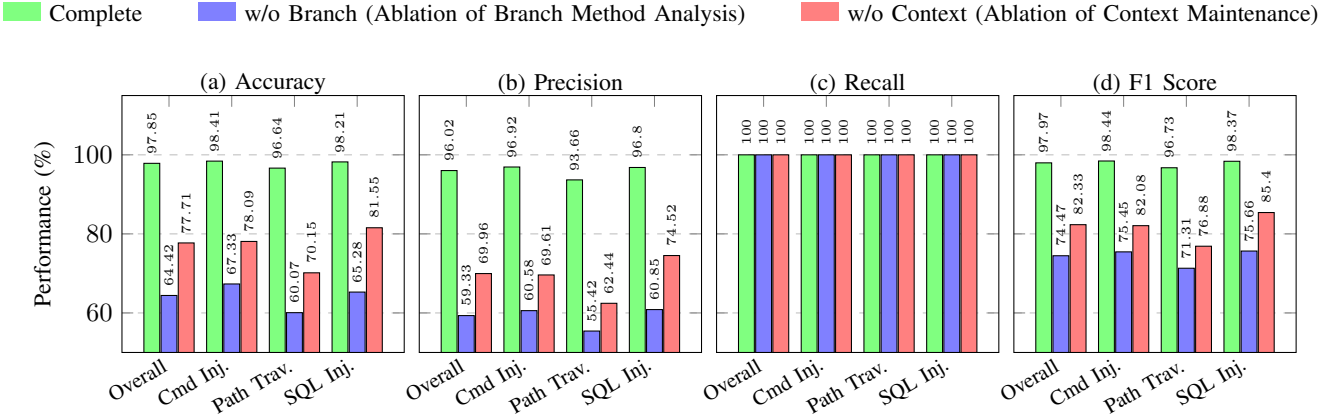
\begin{figure*}[t]
  \centering
  \caption{Ablation Experiment Results Comparison}
  \label{fig:ablation-results}
  
  \begin{minipage}{\textwidth}
    \centering
    \small
    \tikz\fill[green!50] (0,0) rectangle (0.5,0.25); Complete \qquad
    \tikz\fill[blue!50] (0,0) rectangle (0.5,0.25); w/o Branch (Ablation of Branch Method Analysis) \qquad
    \tikz\fill[red!50] (0,0) rectangle (0.5,0.25); w/o Context (Ablation of Context Maintenance)
  \end{minipage}
  
  \vspace{0.5cm}
  
  \begin{tikzpicture}
    \begin{groupplot}[
      group style={
        group size=4 by 1,
        horizontal sep=0.2cm,
        ylabels at=edge left,
        yticklabels at=edge left,
      },
      width=0.3\textwidth,
      height=5cm,
      ymin=50, ymax=115,
      ymajorgrids=true,
      grid style=dashed,
      symbolic x coords={Overall, Cmd Inj., Path Trav., SQL Inj.},
      xtick=data,
      x tick label style={rotate=30, anchor=east, font=\footnotesize, yshift=-3pt},
      enlarge x limits=0.25,
    ]
    
    \nextgroupplot[
      ylabel={Performance (\%)},
      ylabel style={font=\small},
      title={(a) Accuracy},
      title style={yshift=-2ex, font=\small},
    ]
    \addplot[ybar, bar width=6pt, fill=green!50, bar shift=-6.5pt,
             nodes near coords, 
             every node near coord/.append style={font=\tiny, anchor=west, xshift=-6.5pt, rotate=90}
    ] coordinates {
      (Overall,97.85) (Cmd Inj.,98.41) 
      (Path Trav.,96.64) (SQL Inj.,98.21)
    };
    \addplot[ybar, bar width=6pt, fill=blue!50, bar shift=0pt,
             nodes near coords,
             every node near coord/.append style={font=\tiny, anchor=west, rotate=90}
    ] coordinates {
      (Overall,64.42) (Cmd Inj.,67.33) 
      (Path Trav.,60.07) (SQL Inj.,65.28)
    };
    \addplot[ybar, bar width=6pt, fill=red!50, bar shift=6.5pt,
             nodes near coords,
             every node near coord/.append style={font=\tiny, anchor=west, xshift=6.5pt, rotate=90}
    ] coordinates {
      (Overall,77.71) (Cmd Inj.,78.09) 
      (Path Trav.,70.15) (SQL Inj.,81.55)
    };
    
    \nextgroupplot[
      title={(b) Precision},
      title style={yshift=-2ex, font=\small},
    ]
    \addplot[ybar, bar width=6pt, fill=green!50, bar shift=-6.5pt,
             nodes near coords,
             every node near coord/.append style={font=\tiny, anchor=west, xshift=-6.5pt, rotate=90}
    ] coordinates {
      (Overall,96.02) (Cmd Inj.,96.92) 
      (Path Trav.,93.66) (SQL Inj.,96.80)
    };
    \addplot[ybar, bar width=6pt, fill=blue!50, bar shift=0pt,
             nodes near coords,
             every node near coord/.append style={font=\tiny, anchor=west, rotate=90}
    ] coordinates {
      (Overall,59.33) (Cmd Inj.,60.58) 
      (Path Trav.,55.42) (SQL Inj.,60.85)
    };
    \addplot[ybar, bar width=6pt, fill=red!50, bar shift=6.5pt,
             nodes near coords,
             every node near coord/.append style={font=\tiny, anchor=west, xshift=6.5pt, rotate=90}
    ] coordinates {
      (Overall,69.96) (Cmd Inj.,69.61) 
      (Path Trav.,62.44) (SQL Inj.,74.52)
    };
    
    \nextgroupplot[
      title={(c) Recall},
      title style={yshift=-2ex, font=\small},
    ]
    \addplot[ybar, bar width=6pt, fill=green!50, bar shift=-6.5pt,
             nodes near coords,
             every node near coord/.append style={font=\tiny, anchor=west, xshift=-6.5pt, rotate=90}
    ] coordinates {
      (Overall,100) (Cmd Inj.,100) 
      (Path Trav.,100) (SQL Inj.,100)
    };
    \addplot[ybar, bar width=6pt, fill=blue!50, bar shift=0pt,
             nodes near coords,
             every node near coord/.append style={font=\tiny, anchor=west, rotate=90}
    ] coordinates {
      (Overall,100) (Cmd Inj.,100) 
      (Path Trav.,100) (SQL Inj.,100)
    };
    \addplot[ybar, bar width=6pt, fill=red!50, bar shift=6.5pt,
             nodes near coords,
             every node near coord/.append style={font=\tiny, anchor=west, xshift=6.5pt, rotate=90}
    ] coordinates {
      (Overall,100) (Cmd Inj.,100) 
      (Path Trav.,100) (SQL Inj.,100)
    };
    
    \nextgroupplot[
      title={(d) F1 Score},
      title style={yshift=-2ex, font=\small},
    ]
    \addplot[ybar, bar width=6pt, fill=green!50, bar shift=-6.5pt,
             nodes near coords,
             every node near coord/.append style={font=\tiny, anchor=west, xshift=-6.5pt, rotate=90}
    ] coordinates {
      (Overall,97.97) (Cmd Inj.,98.44) 
      (Path Trav.,96.73) (SQL Inj.,98.37)
    };
    \addplot[ybar, bar width=6pt, fill=blue!50, bar shift=0pt,
             nodes near coords,
             every node near coord/.append style={font=\tiny, anchor=west, rotate=90}
    ] coordinates {
      (Overall,74.47) (Cmd Inj.,75.45) 
      (Path Trav.,71.31) (SQL Inj.,75.66)
    };
    \addplot[ybar, bar width=6pt, fill=red!50, bar shift=6.5pt,
             nodes near coords,
             every node near coord/.append style={font=\tiny, anchor=west, xshift=6.5pt, rotate=90}
    ] coordinates {
      (Overall,82.33) (Cmd Inj.,82.08) 
      (Path Trav.,76.88) (SQL Inj.,85.40)
    };
    
    \end{groupplot}
  \end{tikzpicture}
\end{figure*}

Table~\ref{tab:comprehensive-comparison} presents the performance comparison across all three vulnerability types. We use \sname with Kimi-K2 as the primary configuration due to its best overall performance. The results show \sname substantially outperforms all baselines, with accuracy improvements of at least 24\%.

CodeQL achieves 52-57\% accuracy with 50-54\% precision, meaning approximately half of its detections are false alarms requiring manual verification. While maintaining relatively high recall in some categories, this low precision creates substantial workload for security analysts. \sname addresses this by achieving 93.66-96.92\% precision with 100\% recall, effectively reducing false positives without sacrificing detection completeness.

The performance gap against LLM-based baselines is equally significant. CWE-DF achieves 48-52\% accuracy, suggesting limited effectiveness despite using advanced LLMs. Best-Rated Agent performs better at 67.8-74.3\% accuracy but still falls short by at least 24\%. In contrast, \sname demonstrates balanced performance across all metrics, with F1-scores ranging from 96.73\% to 98.44\%.

Different LLM backbones within \sname show varying performance across vulnerability types. Kimi-K2 achieves the highest accuracy for command injection (98.41\%) and SQL injection (98.21\%), outperforming DeepSeek-V3 (93.63\%) and GPT-4o (90.44\%). However, DeepSeek-V3 excels at path traversal with 98.88\% accuracy, surpassing Kimi-K2's 96.64\%. These results indicate different LLMs may have varying capabilities in detecting specific vulnerability patterns.

By maintaining 100\% recall with precision above 93.66\%, \sname achieves comprehensive vulnerability coverage with practical false positive control.

\subsection{RQ2: Ablation Study}
To validate the effectiveness of key components in our method, we designed systematic ablation experiments. The experiments use the Kimi-K2 model that performed best in the complete standard method, quantifying the contribution of each component by removing specific parts.

We designed two ablation experiments to evaluate the contribution of each component. \textit{Ablation of Branch Method Analysis} retains core components such as critical type design, non-exploitable condition design, and progressive analysis on the main path, but removes the semantic extraction logic of branch methods. This experiment aims to verify the role of Branch Method Analysis in improving detection comprehensiveness. \textit{Ablation of Context Maintenance} retains the SAST-based main path extraction method and semantic extraction logic of branch methods, but does not maintain context information (does not progressively maintain critical parameters, parameter states, etc.), directly providing all code to the large model for vulnerability detection. This experiment aims to verify the importance of the Context Maintenance mechanism.

\subsubsection{Ablation Experiment Results}

As shown in Figure~\ref{fig:ablation-results}, the ablation study demonstrates the critical contributions of our method's key components. The uniform 100\% recall across all configurations reveals an important characteristic: LLMs inherently tend to identify potential vulnerabilities in code, but struggle to recognize factors that render these vulnerabilities non-exploitable. This limitation manifests in the substantial precision gaps observed when key components are removed.

The Branch Method Analysis component proves essential for vulnerability assessment. Its removal causes dramatic performance degradation across all metrics, with accuracy plummeting by over 33\% and F1-score declining by more than 23\%. More critically, precision drops from 96.02\% to merely 59.33\%, indicating that without systematic branch analysis, the model generates excessive false positives by failing to identify execution paths that prevent vulnerability exploitation.

Context Maintenance demonstrates similarly vital importance in enabling accurate vulnerability identification. Ablating this component results in notable performance deterioration, with precision dropping to 69.96\% and F1-score declining to 82.33\%. This degradation is particularly pronounced for Path Traversal vulnerabilities, where precision falls to 62.44\%, underscoring the necessity of progressively tracking contextual information throughout the analysis process.
Together, these components significantly enhance the LLM's capability to distinguish truly exploitable vulnerabilities from vast amounts of superficially suspicious code, transforming inherently unpredictable LLM reasoning into systematic, precise, and context-aware security analysis through deliberate architectural design.

\subsection{RQ3: Real-world Practicality Evaluation}

To validate the practicality and effectiveness of our method in real software projects, we designed a systematic evaluation framework and conducted in-depth security analysis on multiple open-source projects.

We designed systematic real-world validation experiments, selecting three representative large-scale open-source projects as test subjects\cite{PublicCMS,JeeWMS,dreamer_cms_2024}. For each project, we adopted a standardized security audit process: SAST tool-based taint path generation for initial screening, in-depth analysis using the constraint solving framework, and manual verification confirmation to ensure the authenticity and accuracy of discovered vulnerabilities.

\subsubsection{Case Study Results and Real-world Impact}

\begin{lstlisting}[style=jsonstyle,caption={Simplified Command Injection Vulnerability},label={lst:rce-example}, float=t]
// Entry point
void handleRequest(FileProperties props, TemplateData data) {
    CodeGenerator generator = new CodeGenerator(props, data);
    generator.generate();
}

void generate() {
    // Conditional call
    if (this.props.shouldGenerateJSP()) {
        invokeTemplate("template.ftl", "jsp");
    }
}

void invokeTemplate(String templateFile, String type) {
    // Data flows deeper
    generateFile(templateFile, type, this.data);
}

void generateFile(String templateFile, String type, Map data) {
    String path = data.get("entityPackage").toString() + "/" + data.get("entityName").toString();
    Writer out = new FileWriter(path);
    // RCE by writing malicious JSP file
    processTemplate(templateFile, data, out);
    out.close();
}
\end{lstlisting}

Through systematic security auditing across three representative large-scale open-source projects, we discovered 15 real security vulnerabilities, fully validating the practical value and effectiveness of our method:

\textit{Vulnerability Discovery and Distribution}: We successfully identified 15 real security vulnerabilities covering critical types: 6 command injection vulnerabilities, 4 path traversal vulnerabilities, and 5 SQL injection vulnerabilities. This diverse distribution demonstrates the comprehensiveness and effectiveness of our method in detecting different threats.

\textit{Illustrative Case}: Listing~\ref{lst:rce-example} presents a simplified representation of a real command injection vulnerability discovered during our evaluation, condensed from a complex multi-layer call chain. Request parameters propagate through multiple method invocations before reaching the sink where malicious JSP files are written. \sname systematically traces this call chain, analyzing parameter states at each layer while maintaining contextual information about data flow and constraints. This methodical, step-by-step analysis enables \sname to identify the vulnerability through predictable reasoning, demonstrating its effectiveness in handling real-world vulnerabilities with complex propagation paths.

\textit{Security Impact Assessment}: The discovered vulnerabilities underwent professional assessment, with generally high risk levels and estimated CVSS scores ranging from 7.5 to 9.8, classified as high to critical severity. If maliciously exploited, these vulnerabilities could lead to serious consequences such as system control acquisition and sensitive data leakage.

\textit{Community Contributions}: We have submitted detailed vulnerability reports to relevant open-source projects and are collaborating with project maintainers on vulnerability confirmation and remediation. Additionally, we follow standard procedures to submit vulnerability applications to the CVE database, contributing to the open-source community.

\section{Related Works}
\label{related_works}

Recent advances in LLM-based code auditing can be broadly categorized into two classes.
The first class leverages LLMs as auxiliary tools to enhance traditional symbolic analyzers.
For instance, IRIS integrates LLM-generated taint specifications with CodeQL to improve whole-repository vulnerability detection, outperforming CodeQL alone in recall and precision benchmarks~\cite{li2024iris}.
Other hybrid approaches employ fine-tuned code LLMs to reduce false positives by enhancing long-context classification~\cite{shestov2025finetuning}.
Similarly, studies like Devign~\cite{zhou2019devign}, CodeBERT~\cite{feng2020codebert}, and VulZoo~\cite{ruan2024vulzoo} , which offers a large-scale vulnerability intelligence dataset widely used to support LLM-augmented analysis.
However, these methods are limited by their inability to support fine-grained, IDE-level semantic analysis and cross-function propagation.

The second class treats LLMs as interpreters to perform end-to-end vulnerability reasoning.
Approaches such as LLMDFA~\cite{wang2024llmdfa} utilize few-shot prompting and external verification tools to enable dataflow analysis without compilation, achieving significantly higher F1 scores compared to traditional tools.
Autonomous agents like RepoAudit~\cite{guo2025repoaudit} explore code repositories demand-drivenly, validating path constraints and mitigating hallucinations during auditing .
Other benchmarks such as VulDetectBench~\cite{liu2024vuldetectbench} , JITVUL~\cite{yildiz2025benchmarking} , and VulnSage~\cite{zibaeirad2025reasoning} evaluate LLMs and ReAct Agents on multi-CVE, repository-level tasks, highlighting the importance of interprocedural context. Recent work has also extended LLM reasoning to smart contracts, notably GPTScan~\cite{sun2024gptscan}, which combines GPT with static program analysis to detect logic vulnerabilities.

Despite their strengths, these methods still face critical limitations: reasoning often remains function-local (lacking structured propagation across call chains), control/data-flow constraints are implicitly modeled, and exploitability depends on compound interprocedural logic that existing agents struggle to capture .
Hybrid approaches like LLMVulExp~\cite{mao2025towards}  demonstrate potential improvements in explainability and evaluation, but robustness under industrial-scale workloads remains an open challenge.

Our work builds on these insights by introducing \sname, a unified LLM+SAST framework that explicitly models source-to-sink paths as sequences of subtasks.
By defining critical types and unexploitable conditions, and maintaining dynamic context across method calls, \sname enables human-like, progressive reasoning.
Unlike RepoAudit, \sname provides both interpretable constraint flow and fine-grained analysis, significantly improving precision and robustness in complex vulnerability scenarios.

\section{Conclusion}


In this work, we formalize vulnerability discovery as a constraint-solving problem, introducing Transfer Constraints and Trigger Constraints to determine whether a call chain contains exploitable vulnerabilities.
To manage the complexity of constraint solving, we decompose the task into subtasks along the call chain, each analyzing method pairs and maintaining context until the Trigger Constraints at the sink can be verified—enabling focused analysis on essential code while distilling complex context into conclusions. This decomposition strategy significantly reduces the complexity of reasoning while preserving semantic precision across the analysis path.
We implement \sname, a framework that realizes this algorithm, conduct experiments to validate its performance.
Our experiments show that \sname achieves promising performance in vulnerability detection, including 97.85\% accuracy and a 97.97\% F1-score on OWASP benchmarks.
Moreover, \sname uncovered 15 previously unknown high-severity vulnerabilities in real-world projects, demonstrating its practical effectiveness.


\clearpage
\bibliographystyle{IEEEtran}
\bibliography{contents/references}

\end{document}